\documentclass[english,twocolumn,amsmath,amssymb]{revtex4}

\usepackage[T1]{fontenc}
\usepackage[latin1]{inputenc}
\usepackage{graphicx}
\usepackage{amsmath}
\usepackage{epstopdf}
\makeatletter

\usepackage{bm}

\usepackage{babel}
\makeatother

\begin{document}

\title{Radio-frequency capacitance spectroscopy of metallic nanoparticles}

\author{J.C. Frake,$^{1}$ S. Kano,$^{2,3}$ C. Ciccarelli,$^{1}$ J. Griffiths,$^{1}$ M. Sakamoto,$^{3,4,5}$ T. Teranishi,$^{3,4}$ Y. Majima,$^{2,3,6}$ C. G. Smith,$^{1}$ and M. R. Buitelaar$^{7,8}$}

\affiliation{\vspace{1.5em}
$^{1}$Cavendish Laboratory, University of Cambridge, Cambridge CB3 0HE, United Kingdom\\
$^2$Materials and Structures Laboratory, Tokyo Institute of Technology, Yokohama 226-8503, Japan\\
$^3$CREST, Japan Science and Technology Agency, Yokohama 226-8503, Japan\\
$^4$Institute for Chemical Research, Kyoto University, Uji 611-0011, Japan\\
$^5$PRESTO, Japan Science and Technology Agency, Uji 611-0011, Japan\\
$^6$Department of Printed Electronics Engineering, Sunchon National University, Sunchon 540-742, South Korea\\
$^7$London Centre for Nanotechnology, UCL, London WC1H 0AH, United Kingdom\\
$^8$Department of Physics and Astronomy, UCL, London WC1E 6BT, United Kingdom \bigskip}

\begin{abstract}
Recent years have seen great progress in our understanding of the electronic properties of nanomaterials in which at least one dimension measures less than 100 nm. However, contacting true nanometer scale materials such as individual molecules or nanoparticles remains a challenge as even state-of-the-art nanofabrication techniques such as electron-beam lithography have a resolution of a few nm at best. Here we present a fabrication and measurement technique that allows high sensitivity and high bandwidth readout of discrete quantum states of metallic nanoparticles which does not require nm resolution or precision. This is achieved by coupling the nanoparticles to resonant electrical circuits and measurement of the phase of a reflected radio-frequency signal. This requires only a single tunnel contact to the nanoparticles thus simplifying device fabrication and improving yield and reliability. The technique is demonstrated by measurements on 2.7 nm thiol coated gold nanoparticles which are shown to be in excellent quantitative agreement with theory.
\end{abstract}




\maketitle

Electrical characterization of metallic nanoparticles using dc transport methods requires both a source and drain electrode to pass a current as well as a gate electrode to vary the chemical potential, see Fig. 1a. The most commonly used fabrication techniques are electromigration, where a metallic wire is controllably loaded to failure such that a nanosize gap opens up in which particles might be trapped, or electrode plating in which a predefined gap is reduced electrochemically \cite{Li1,Kuemmeth,Azuma}. Although these methods are certainly feasible, the need for two tunnel contacts makes the fabrication complex and challenging and the quality and yield of devices highly variable. This is particularly problematic for the smallest nanoparticles - or molecules - less than around 3 nm in size.

Here we present a different technique that allows the study of arbitrarily small nanoparticles using only a single tunnel contact. Importantly, the technique does not require nm positioning accuracy which facilitates the use of standard lithography methods. As illustrated in Fig. 1b, we achieve this by placing the nanoparticles on a metal electrode and and by \textit{capacitively} coupling the nanoparticle to a second electrode, which functions both as an ac reference ground and as a dc gate electrode. The devices are embedded in a resonant LC circuit and radio-frequency reflectometry is used as the measurement tool \cite{Duty, Petersson, Chorley, Verduijn}. This approach, which can be considered as capacitance spectroscopy \cite{Ashoori1,Ashoori2} taken into the radio-frequency domain, not only simplifies device fabrication but also offers high sensitivity and high bandwidth readout.

\begin{figure*}
\includegraphics[width=170mm]{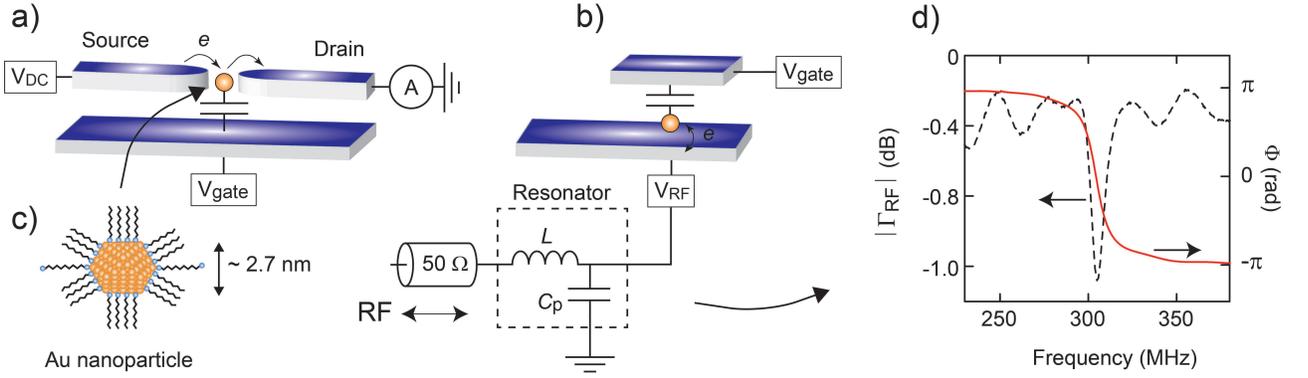}
\caption{\label{Fig1} \textbf{(a)} Schematic of a conventional direct-current measurement. Two transport electrodes that are precisely aligned and separated by a few nanometers are required to enable electron transport through a nanoparticle. \textbf{(b)} Schematic of the radio-frequency reflectometry technique. The nanoparticle is only tunnel coupled to a \textit{single} electrode. A second electrode, which does not need to be precisely aligned, functions both as an ac ground and dc gate electrode. The device is embedded in a LC resonant circuit which both simplifies device fabrication and allows for high sensitivity and high bandwidth measurements. \textbf{(c)} Representation of a gold nanoparticle coated in thiol chains used in our experiments. \textbf{(d)} Measured amplitude and phase response of a resonant circuit with resonant frequency $ \sim $ 305 MHz.}
\end{figure*}

The resonant circuits used in our experiments typically have a resonant frequency $f_0$ in the 300-500 MHz range which is given by the total device capacitance $C_{\Sigma}$ - including parasitics - and a chip inductance $L$ placed on the sample holder such that $f_0 = 1/(2\pi \sqrt{LC_{\Sigma}})$, see Fig. 1d. The key idea is that the capacitance depends in part on the ability of electrons to move on and off the nanoparticle at the driving frequency which in turn depends on the chemical potential of the nanoparticle which can be tuned by the gate electrode. The gate-dependent capacitance is probed by measuring the phase shift of a reflected radio-frequency signal at the resonant frequency $f_0$.

Before discussing experimental data, using Au nanoparticles, we will first examine, quantitatively, the expected phase response of the reflectometry technique for a generic 'electron-in-a-box' system or quantum dot with discrete energy spectrum, coupled to a metallic lead at some finite temperature $T$, as illustrated in Fig. 2. In the presence of a radio-frequency signal, as in Fig. 1b, the energy levels on the quantum dot will oscillate with respect to the lead. However, it is only possible for electrons to move on and off the quantum dot when an energy level is aligned with the thermally broadened electrochemical potential of the lead. In this case, the occupation probability of the level varies with the drive frequency and the movement of charge on and off the electrode in response to the rf drive amplitude can be parameterized by an effective admittance $Y = 1/\Delta R + j \omega \Delta C$ where $\omega = 2 \pi f$ is the radial frequency and with capacitance $\Delta C$ and resistance $\Delta R$ as shown in Fig. 2b. If the thermal energy is larger than the lifetime broadening of the resonances, $k_B T \gg h \gamma$, where $\gamma$ is the tunnel rate, it is relatively straightforward, see supplementary information, to obtain the following relations using rate equations:

\begin{equation}\label{Reff}
 \Delta R = \frac{4 k_B T}{e^2 \alpha^2 \gamma}\left(\frac{\gamma^2}{\omega^2}+1\right) \cosh^{2}\left(\frac{-e \alpha \Delta V_g}{2 k_B T}\right)
\end{equation}

\begin{equation}\label{Ceff}
\Delta C = \frac{e^2 \alpha^2}{4 k_B T}\left(\frac{\omega^2}{\gamma^2}+1\right)^{-1} \cosh^{-2}\left(\frac{-e \alpha \Delta V_g}{2 k_B T}\right)
\end{equation}

\noindent where the term $-e \alpha \Delta V_g$ takes into account the position of the discrete energy level with respect to the Fermi level of the lead using a lever arm $\alpha$ for conversion from gate voltage to energy \cite{Chorley, Gabelli}. If the tunnel rate is comparable to the drive frequency, $\gamma \sim \omega$, dissipation can be significant as previously observed in single-electron tunneling devices \cite{Persson, Ciccarelli}. However, if tunnel rates are considerably faster than the drive frequency, that is $\gamma \gg \omega$, the effective resistance $\Delta R$ diverges and the junction will be capacitive only. It is this limit that we will consider here. Using standard network analysis \cite{Pozar}, it can be shown, see supplementary information, that in this case the measured phase shift, for an under-coupled resonator, is related to the effective capacitance as:

\begin{equation}\label{delta}
\Delta \Phi = - 2 Q \Delta C / C_{\Sigma}.
\end{equation}

\noindent where $Q$ is the quality factor of the resonator given by the ratio of the resonant frequency and bandwidth $\Delta f$ of the resonance. Using the relations 1-3 above we now have the tools to quantitatively investigate single-electron tunneling of any device of nm dimensions in which the energy spectrum is discrete.

\bigskip
\centerline{\large \textbf{Results}}
\medskip

\noindent \textbf{Nanoparticle measurements}. To demonstrate the technique experimentally, we have deposited thiol-coated Au nanoparticles of 2.7 nm diameter on a gold substrate. These particles were then covered by approximately 5.5 nm of Al$_2$O$_3$ deposited using atomic layer deposition, see Methods. On top of the oxide layer we defined a set of gate electrodes of varying sizes as illustrated by the scanning electron microscope (SEM) images in Fig. 3a. The results of measurements on two representative structures of different size in a He-3 cryostat with a base temperature of $\sim$ 400 mK are shown in Fig. 3b. The number of particles under the large 5 x 5 $\mu$m gate electrodes is not exactly known but, given an approximate density of Au nanoparticles of $\sim$ 10-50 $\mu$m$^{-2}$, see supplementary information, is likely to be large. Therefore, even though for individual particles the energy levels are expected to be well separated, the measurements on this gate electrode show an interleaving of the signal of many particles, resulting in the oscillatory pattern seen in the right panel image of Fig.~3b. For the measurements shown in the left panel of Fig.~3b, the gate overlap is significantly reduced - by a factor of 25 -  and we observe more pronounced dips against a relatively flat background. For these devices the number of nanoparticles under the gate electrodes is sufficiently low for the observation of tunneling through individual electron states as we will examine in more detail. We also performed a series of control experiments, see supplementary information, in which no nanoparticles were deposited and in which we did not see any features in the measurements.

\begin{figure*}
\includegraphics[width=140mm]{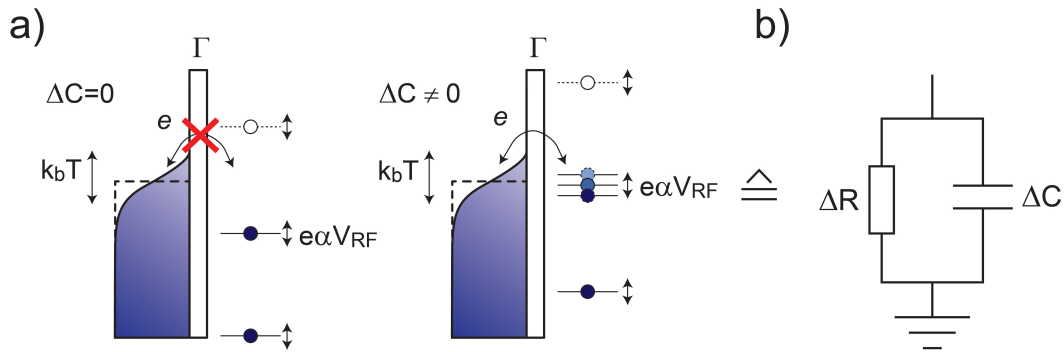}
\caption{\label{Fig2} \textbf{(a)} Schematic energy diagrams showing the discrete energy states of the nanoparticle and the temperature broadened electron distribution in the lead. The occupation probabilities of the states on the nanoparticle depend on the relative position of the states with respect to the electrochemical potential of the lead and varies from fully occupied (filled circles) to unoccupied (open circles). In response to the rf excitation, the relatively position of the levels oscillate at the applied frequency. Left: without a state in the thermal window around the electrochemical potential of the lead, no tunneling is allowed: all states are either fully occupied or empty. Right: when a level is aligned with the electrochemical potential of the lead, the average occupancy of the state oscillates at the driving frequency and charge thus moves back and forth between the particle and the lead \textbf{(b)} The device can be parameterized by an effective RC circuit as explained in the main text.}
\end{figure*}

Using the relations 2 and 3 obtained above for $\Delta \Phi$ versus $\Delta V_g$ we are able to compare the experimental data with theory. We obtain an excellent fit for individual resonances, see Fig.~3c, where the only free parameter is the lever arm $\alpha$ which sets both the width and depth of the dips. The width of the dips $\Delta V_g \propto T/\alpha$, or, more precisely, the full-width-half-maximum is 3.53 $k_B T/e \alpha$ which is identical to what is expected in dc transport of individual quantum states \cite{Meirav, Beenakker, Foxman}. However, unlike dc transport in which the magnitude of the conductance is set by the tunnel rate and asymmetry of the source and drain tunnel junctions, in the reflectometry measurements - in which there is only a single tunnel junction - the magnitude $\Delta \Phi \propto \alpha^2 / T$. We thus have two relations for $\alpha$ and $T$ which can be obtained independently. Indeed, while in our experiments the temperature is known, $T \sim 400$ mK, if the temperature is set as an additional fit parameter, the fit procedure yields the correct experimental result. This enables an energy calibration without the need of a source-drain bias voltage and allows measurements of, for example, the charging energy of the nanoparticles or, when a magnetic field is applied, measurements of the g-factor or energy spectrum \cite{Ashoori1, Ashoori2}.

\begin{figure*}
\includegraphics[width=115mm]{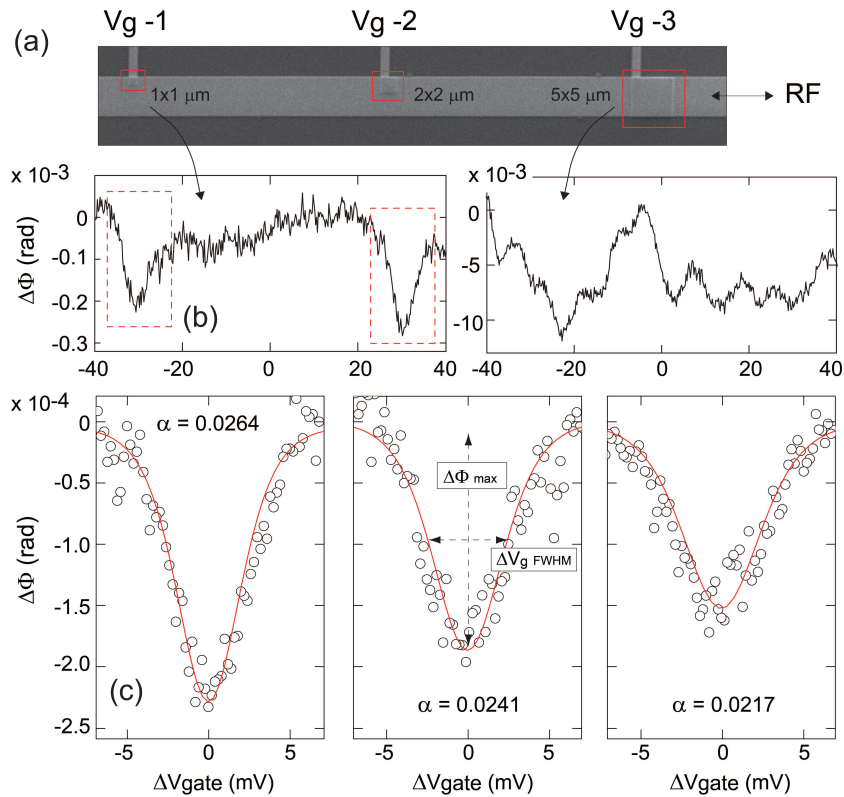}
\caption{\label{Fig3} \textbf{(a)} Scanning electron microscope image of a nanoparticle device showing three gate electrodes with different overlap with the rf electrode. \textbf{(b)} Gate dependent phase response for a 1x1 $\mu$m (left) and 5x5 $\mu$m gate area (right). For the smallest gate areas fewer particles are being detected and individual resonances are observed. \textbf{(c)} Fitted curves to three individual dips observed in a small device. A constant background slope seen in all data, including control samples, has been subtracted. For the dips we obtain $\alpha = 0.0264, 0.0241,$ and 0.0217 from left to right, respectively. Since the widths of the dips scale, for a given temperature $T$, as $\Delta V_{g~\textrm{FWHM}} \propto 1/ \alpha$ and their depths as $\Delta \Phi_{\textrm{max}} \propto \alpha^2$ the dips with lower $\alpha$ are both wider and shallower.}
\end{figure*}

\noindent \textbf{Nano-thermometry}. A further important consequence is that our devices can be used as a primary thermometer, that is, a thermometer that allows temperature measurements without calibration by another thermometer. The advantage of using a thermometer based on nanoparticles is that it works over an exceptionally large temperature range. The lower limit of the thermometer is set by the requirement that lifetime broadening due to the coupling to the lead is less than thermal smearing: $2 h \gamma < k_B T$. Since at the same time, the tunnel rates should be comparable to or larger than the rf driving frequency, a practical limit would be $\gamma \gtrsim 100$ MHz which corresponds to about 10 mK. The higher limit of the thermometer is set by the spacing between adjacent dips in measurements as in Fig. 3b. For a single nanoparticle this separation is given by the addition energy which exceeds several tens of meV \cite{Kano2} for small nanoparticles and which corresponds to room temperature operation \cite{Kano1}. The energy-level separation $\Delta E$ of Au nanoparticles of 2.7 nm diameter is of order 10 meV such that for temperatures above around 100 K a transition from the quantum ($\Delta E \gg k_B T$) to the classical ($\Delta E \ll k_B T$) transport regime is expected. In this case Eq. 2 is no longer valid, although the resonances will still be thermally broadened and can be modeled \cite{Beenakker, Foxman}. A further practical consideration is that the signal strength $\Delta \Phi \propto \alpha^2 / T$, and thus decreases with increasing temperature. This can be compensated for by increasing the lever arm $\alpha$ which is set by the gate oxide thickness and dielectric constant. Using Al$_2$O$_3$, we found $\sim$ 5 nm to be the optimal oxide thickness for our devices: thicker oxides reduced the signal-to-noise ratio while thinner oxides affected device yield and resulted in breakthrough when the gate voltage was increased beyond $\sim 0.5$ V. Nevertheless, the lever arm can be increased significantly by using high-dielectric oxides such as HfO$_2$ or TiO$_2$ which would improve signal-to-noise ratios by about two orders of magnitude. By a further improvement of the quality factor of the resonator, for example by reducing parasitic capacitances, we believe that room temperature operation is feasible. By embedding the device in a resonant circuit, as in this work, measurements are also fast and not limited by 1/f noise as in conventional dc quantum dot thermometry \cite{Pekola, Maradan}.

\bigskip
\centerline{\large \textbf{Discussion}}
\medskip

The experiments on Au nanoparticles described here are readily extended to other types of nanoparticles such as Pt, Fe, or Co nanoparticles, which have applications in a range of areas varying from catalysis to biomedicine \cite{Schmid,Lu}, single molecule magnets such as Mn$_{12}$, Cr$_7$Ni or TbPc$_2$, or endohedral fullerenes which are of interest in spintronics or spin-based quantum information processing \cite{Ardavan, Thiele, Brown}. As identified above, there is furthermore significant scope for improvement in the measurement sensitivity - and thus the operating temperature - of the nanodevices by optimization of the oxide dielectrics and quality factor of the resonators used. Chemical modification of the length of linker molecules, such as decanedithiol used in this work, provides control over the tunnel rates while varying the nanoparticle solution pH and adsorption time \cite{Zhu}, and surface modification of the electrodes \cite{Bhat,Zirbs,Li2} - or a combination of these - allows nanoparticle density control. Taken together, this opens up novel routes for the characterization and application of individual molecules and metallic nanoparticles.

\bigskip
\centerline{\large \textbf{Methods}}
\medskip

\noindent Devices were fabricated on nominally undoped Si wafer substrates of resistivity > 100 Ohm cm patterned with Ti/Au electrodes using electron-beam lithography. Before deposition of the nanoparticles, the electrodes were cleaned using acetone, ethanol and an oxygen plasma, followed by a further ethanol immersion to remove any surface oxides. The devices were then immersed in a toluene-based gold nanoparticle solution for one hour, followed by a rinse with toluene and ethanol. The gold nanoparticles have a 2.7 nm core diameter and were coated with octanethiol (length: 1.44 nm) and decanedithiol (1.82 nm) mixed monolayer. The devices were subsequently covered by approximately 5.5 nm of Al$_2$O$_3$ using atomic layer deposition. In the final stage, a further set of Ti/Au gate electrodes were defined on the devices using electron-beam lithography. The devices typically showed no leakage between electrodes separated by the Al$_2$O$_3$ layer up to around 0.5 V, at which point the oxide coating on the devices began to break down.
\newline
\newline
\noindent The devices were mounted on a microstrip line coated printed circuit board sample holder with radio-frequency connections. Measurements were carried out at $T \sim 400$ mK in a Helium-3 cryostat, custom modified to allow reflectometry radio-frequency measurements, with a low noise cryogenic amplifier to improve the signal-to-noise ratio of the reflected signal.  Demodulation is achieved by mixing the reflected rf signal with the reference signal. Both quadratures of the signal are detected, allowing measurements of both the amplitude and phase response. For the measurements shown in Fig.~3 a chip inductor $L = 500$ nH was used. This yielded a resonance frequency $f_0 = 345$ MHz and capacitance $C_\Sigma = 0.43$ pF with quality factor Q $\sim 59$  which were the parameters used in the fit procedure for Fig.~3c. The dc gate voltage was applied to the samples via a bias-tee.

\bigskip
\centerline{\large \textbf{Author contributions}}
\medskip

\noindent J.C.F. fabricated the samples, set up the measurement system and performed the measurements under supervision of C.G.S and M.R.B who designed the experiments. M.S. and T.T. synthesized the gold nanoparticles which were deposited on the substrates by S.K under supervision of Y.M. Atomic layer deposition was performed by C.C. Electron-beam lithography was performed by J.G. Together J.C.F and M.R.B. analyzed the data and wrote the manuscript on which all authors commented.

\onecolumngrid
\newpage

\setcounter{page}{1} \thispagestyle{empty}

\begin{center}
\textbf{{\large Radio-frequency capacitance spectroscopy of metallic nanoparticles:\\Supplementary Information}}\\
\bigskip
J.C. Frake,$^{1}$ S. Kano,$^{2,3}$ C. Ciccarelli,$^{1}$ J. Griffiths,$^{1}$ M. Sakamoto, $^{3,4,5}$ T. Teranishi, $^{3,4}$ Y. Majima,$^{2,3,6}$ C. G. Smith,$^{1}$ and M. R. Buitelaar$^{7,8}$

\bigskip

\textit{$^{1}$Cavendish Laboratory, University of Cambridge, Cambridge CB3 0HE, United Kingdom\\
$^2$Materials and Structures Laboratory, Tokyo Institute of Technology, Yokohama 226-8503, Japan\\
$^3$CREST, Japan Science and Technology Agency, Yokohama 226-8503, Japan\\
$^4$Institute for Chemical Research, Kyoto University, Uji 611-0011, Japan\\
$^5$PRESTO, Japan Science and Technology Agency, Uji 611-0011, Japan\\
$^6$Department of Printed Electronics Engineering, Sunchon National University, Sunchon 540-742, South Korea\\
$^7$London Centre for Nanotechnology, UCL, London WC1H 0AH, United Kingdom\\
$^8$Department of Physics and Astronomy, UCL, London WC1E 6BT, United Kingdom}

\end{center}

\setcounter{figure}{0}
\setcounter{equation}{0}

\renewcommand{\figurename}{Figure S}

\noindent The following sections comprise the supplementary material to the main text. We provide the following: (i) An analysis of the RLC network used in the experiments; (ii) A derivation of the effective admittance of the nanoparticle devices, starting with the rate equations and (iii) The procedure used for the data fits and control experiments data in which no Au nanoparticles were present.


\section{RLC Network}

\noindent A diagram of the circuit used in this work is shown in Fig. S1. The circuit consists of a capacitance $C_p$ which takes into account the parasitics of the components and wiring, an inductor $L$ which consists of the chip inductor placed on the sample holder, and a resistor $R_m$ which takes into account the resistances in the wiring and the inductor - but could also double to allow for a matching resistor. The nanoparticle device, when modeled as a lumped component resistor $\Delta R$ and capacitor $\Delta C$ in parallel, as described in Section II below, allows the treatment of the entire system as a simple tank circuit. The impedance of the resonant circuit can then be described as:

\begin{figure*}[b]
\includegraphics[width=120mm]{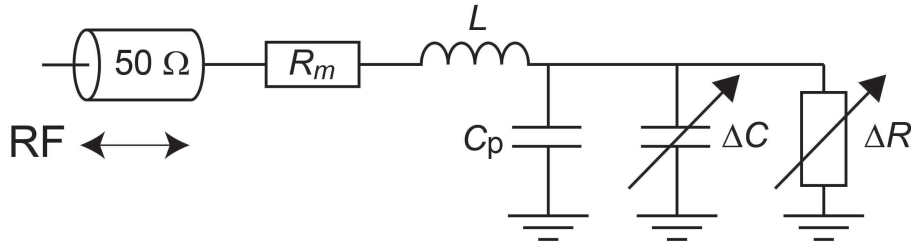}
\caption{\label{FigS1} Schematic model of the RLC circuit used in the radio-frequency reflectometry measurements. The nanoparticle device is modeled as a lumped component resistor $\Delta R$ and capacitor $\Delta C$ in parallel which allows the treatment of the entire system as a simple tank circuit.}
\end{figure*}

\begin{equation}
 Z_{tot}=R_{m}+i\omega L+ \Delta R \mathbin{\parallel} (\Delta C + C_p)
\end{equation}
\\
\noindent Expanding out the terms and arranging into real and imaginary components, we arrive at the impedance of the circuit as a function of the `components' of the device, $\Delta R$ and $\Delta C$.

\begin{equation}
 Z_{tot}=\left[R_{m}+\frac{\Delta R}{1+(\omega C_{\Sigma} \Delta R)^2}\right] + i\omega \left[L-\frac{C_{\Sigma} {\Delta R}^{2}}{1+(\omega C_{\Sigma} \Delta R)^2}\right]
\end{equation}
\\
\noindent where $C_{\Sigma} = C_{p} + \Delta C$.  From here, it is possible to understand how a shift in the resistance or capacitance of the device will change the response seen on the circuit. We can simplify this equation by looking at how it behaves near resonance, since this is the nominal operation range - as carried out in the resonant circuit model of Ref.~\cite{Roschier}. Making the (generally good) approximation that $ \omega \Delta R C_{\Sigma} >> 1$ and substituting $\omega_0^{2}= 1 / (LC_{\Sigma})$:

\begin{equation}
Z_{tot, res}=R_{m}+\frac{L}{\Delta R C_{\Sigma}}+i\omega L-\frac{i}{\omega C_{\Sigma}}
\end{equation}
\\
\noindent where the imaginary component vanishes to zero exactly on resonance. Note that in our work $\Delta R$ is typically very large, which differs from the work by, e.g., Schoelkopf et al \cite{Schoelkopf} where the resonant circuit is used to down-convert the resistance of a single-electron transistor (typically of order 10 k$\Omega$) to 50 $\Omega$. From here on, unless explicitly stated, the frequency is assumed to be on resonance. If we define $Z_{r}$ and $Z_{i}$ as the real and imaginary components of the impedance of the circuit, and $Z_{0}$ as the $50\, \Omega$ impedance of the wave guide connected to the tank circuit, then the reflection coefficient is given by:

\begin{equation}
 \Gamma=\frac{Z-Z_{0}}{Z+Z_{0}}=\frac{Z_{r}^{2}+Z_{i}^{2}-Z_{0}^{2}+2iZ_{0}Z_{i}}{(Z_{r}+Z_{0})^{2}+Z_{i}^{2}}
\end{equation}
\\
\noindent This, along with the calculated impedance from above, can be put together to show how the measured response of the tank circuit will vary as a function of effective dissipative $\Delta R$ and reactive $\Delta C$ components of the device, which can be in turn matched to the physics of the device itself using appropriate models.\\

\noindent The amplitude and phase of the reflected signal are given by:

\begin{align}
|\Gamma| & =\frac{\sqrt{(Z_{r}^2+Z_{i}^2-Z_{0}^2)^2+(2Z_{0}Z_{i})^2}}{(Z_{r}+Z_{0})^2+Z_{i}^2}\\ \nonumber
\\
\Gamma_{\phi} & =\textrm{atan} \frac{2Z_{0}Z_{i}}{|Z|^2-Z_{0}^{2}}
\end{align}
\\
\noindent The phase response is a clear indicator of under [$Z(\omega_{res}) > 50 \Omega$] or over [$Z(\omega_{res}) < 50 \Omega$] coupling.  In the latter case, the phase change through resonance will be $ 0 \leq \phi \leq \pi$ and the former will result in a phase shift of $  \phi = 2\pi$. It is now possible to see how the magnitude and phase of the reflection coefficient vary with relevant factors.  For the capacitance shift in phase of the reflection coefficient:

\begin{equation}
 \left. \frac{\partial \Gamma_{\phi}}{\partial \Delta C} \right|_{\omega = \omega_{0}} = \frac{2 Z_{0}}{\omega C_{\Sigma}^{2}(R_{eff}^{2}-Z_{0}^{2}) } = \frac{2 Q Z_{0}}{C_{\Sigma} (R_{eff}-Z_{0})},
\end{equation}
\\
\noindent where $R_{eff}=R_{m}+ L/ (\Delta R C_{\Sigma})$ and $Q = \sqrt{\frac{L}{C_{\Sigma}}} /(R_{eff} + Z_0)$ is the loaded quality factor of the resonator. In the limit where $R_m \rightarrow 0$ and $\Delta R \rightarrow \infty$ this reduces to Eq. 3 in the main text. For the change in reflection magnitude due to a device resistance shift (still assuming $ \omega \Delta R C_{\Sigma} >> 1$) :

\begin{equation}
\left.  \frac{\partial |\Gamma|}{\partial \Delta R} \right|_{\omega = \omega_{0}} = \frac{2 L Z_{0}}{C_{\Sigma} {\Delta R}^2 (R_{eff}+Z_{0})^{2}} = \frac{2 Z_{0} Q^{2}}{{\Delta R}^{2}}.
\end{equation}
\\
\noindent And finally, the change in magnitude due to a capacitance shift:

\begin{equation}
\left.  \frac{\partial |\Gamma|}{\partial \Delta C} \right|_{\omega = \omega_{0}} = \frac{-2 Z_{0} Q^{2}}{C_{\Sigma} \Delta R}.
\end{equation}
\\
\noindent From this, it can be seen that for a fixed device capacitance, the magnitude shift becomes far more sensitive to resistive changes for a device with smaller resistance. It should also be said that the change of reflection coefficient angle is negligible for changes of device resistance in the case where $\Delta R$ is large.

\section{Effective admittance: rate equations}

\noindent To calculate the effective admittance of the nanoparticle device we start by assuming that we are in the incoherent regime such that the tunnel coupling $\gamma$ is sufficiently small: $h \gamma \ll k_b T$. The tunnel rates for a single electron to tunnel off and on a level at energy $\epsilon$ relative to the Fermi energy of the tunnel electrode are given by \cite{MacLean, Petersson2}:

\begin{align}
\gamma_{off} & = \gamma (1-f(\epsilon)) \\
\gamma_{on} & = \eta \gamma f(\epsilon)
\end{align}
\\
\noindent where $\eta =2$ indicates spin degeneracy and $f(\epsilon)$ is a Fermi function. To simplify the analysis we will use $\eta = 1$ here which does not affect the results - apart from a temperature-dependent shift of the resonance center \cite{Bonet} which is not important for the analysis below. In the presence of an rf drive, the position of the level relative to the Fermi energy of the tunnel electrode is given by:

\begin{equation}
\epsilon (t) = - e \alpha (V_g + V_{RF} e^{i \omega t})
\end{equation}
\\
\noindent where we make use of the lever arm $\alpha = C_R/(C_L + C_R)$ where $C_R$ and $C_L$ are the geometric capacitances between the nanoparticle and the gate electrode and the nanoparticle and the rf electrode, respectively. Typically for our devices $C_R \ll C_L$ and as a result $\alpha \ll 1$. The probabilities for the level to be either empty ($P_0$) or occupied ($P_1$) by a single charge are:

\begin{align}
\frac{\mathrm{d} P_{0}}{\mathrm{d} t} & = \gamma_{off} P_1 - \gamma_{on} P_0 \\
\frac{\mathrm{d} P_{1}}{\mathrm{d} t} & = \gamma_{on} P_0 - \gamma_{off} P_1
\end{align}
\\
\noindent Since $P_0 + P_1 = 1$ we arrive at the master rate equation:

\begin{equation}
\frac{\mathrm{d} P_{1}}{\mathrm{d} t}+\gamma P_{1}= \gamma f(\epsilon) \approx \gamma \left(1+\left(1-\frac{e \alpha V_{RF}}{k_B T}\mathrm{e}^{i \omega t}\right) \mathrm{e}^{\frac{-e \alpha V_{g}}{k_B T}}\right)^{-1}
\end{equation}
\\
\noindent where we use a small excitation approximation, i.e. $e \alpha V_{RF} \ll k_B T$.  Notice that the gate voltage term cannot be assumed small in comparison with thermal energies. Using $G= \mathrm{e}^{-(e \alpha V_g)/ (k_B T)}$ for brevity, the solution to this differential equation is a hypergeometric function \cite{Arfken}:

\begin{equation}
P_1= \left( \frac{1}{1+G}\right) \,_2\mathrm{F}_{1} \left( 1,-\frac{i \gamma}{\omega} ; 1-\frac{i \gamma}{\omega};
\frac{G}{1+G} \frac{e \alpha V_{RF} \mathrm{e}^{i \omega t}}{k_B T} \right)
\end{equation}
\\
\noindent This can be expressed as a sum:

\begin{equation}
P_{1}=\frac{1}{1+G} \left[1+\sum_{n=1}^{\infty} \frac{-i \gamma}{\omega n! (n-\frac{i \gamma}{\omega})} \left( \frac{G}{1+G} \frac{e \alpha V_{RF} \mathrm{e}^{i \omega t}}{k_B T} \right )^{n} \right]
\end{equation}
\\
\noindent This is the full response of a single electron revoir with a single lead and single gate to a small oscillation applied to the source.  The terms in this sum die off rapidly with increasing n, so by taking the first term ($n=1$), we can extract much of the physics of the system without too much complexity.

\begin{equation}
P_{1} \approx \frac{1}{1+G}\left[ 1-\frac{\frac{i\gamma}{\omega}}{1-\frac{i\gamma}{\omega}}\frac{G}{1+G}\frac{e \alpha V_{RF} \mathrm{e}^{i\omega t}}{k_B T} \right]
\end{equation}
\\
\noindent The expected charge drawn from a voltage source connected to the RF electrode at time t is given by $Q_{RF} (t) = e \alpha P_1 (t)$. The expected current flowing from the voltage source is then:

\begin{equation}
i_{RF} (t) = \frac{dQ_{RF} (t)}{dt} = \frac{G}{(1+G)^2}\frac{\omega \gamma}{\omega - i \gamma}\frac{e^2 \alpha^2}{k_B T} V_{RF} \mathrm{e}^{i\omega t}
\end{equation}
\\
\noindent The effective impedance is thus:

\begin{equation}
Z_{eff} = V_{RF} e^{i \omega t}/i_{RF} = \frac{(1+G)^2}{G} \frac{k_B T}{e^2 \alpha^2}\frac{\omega^2 + \gamma^2}{\omega \gamma (\omega + i \gamma)}
\end{equation}
\\
\noindent Equating this with $Z_{eff}^{-1} = R_{eff}^{-1} + j \omega C_{eff}$ we find that:

\begin{align}
R_{eff} & =\frac{k_B T}{e^{2}\alpha^2\gamma}\left(\frac{\gamma^{2}}{\omega^{2}}+1 \right)\frac{(1+G)^{2}}{G} \\
C_{eff} & =\frac{e^{2}\alpha^2}{k_B T}\left(\frac{\omega^{2}}{\gamma^{2}}+1 \right)^{-1} \frac{G}{(1+G)^{2}}
\end{align}
\\
\noindent Identifying $(1+G)^2/G$ as $4 \cosh^{2}\left(\frac{-e \alpha \Delta V_g}{2 k_B T}\right)$ we arrive at Eq. 1 and 2 in the main text.

\section{Data analysis and control devices}

\noindent Using Eq.~7 and Eq.~22 (or Eq.~2 and Eq.~3 in the main text) we are able to fit the measured data to theory, provided $C_\Sigma$ and $Q$ are known. We obtain both from the measured amplitude and phase response over a large frequency range which shows a resonance at $f=345$ MHz. Knowing the value of the chip inductor $L=500$ nH this yields $C_\Sigma = 0.43$ pF. The quality factor is measured by fitting the phase response to: $\phi (f) = \phi_0 + 2\> \textrm{atan}[2Q ( 1-f/f_0)]$ where $\phi_0$ is the angle at the resonance frequency $f_0$. This yielded a best fit to the data of $Q=59 \pm 4$. This is somewhat larger than expected for a $Z_0 = 50$ $\Omega$ loaded circuit which we tentatively attribute to an impedance mismatch at the cryogenic amplifier connected to the circuit. For the data fits in the main text we used $\gamma \gg \omega$ and the experimentally determined $C_\Sigma = 0.43$ pF and $Q=59$. A constant background slope, seen in all data including control samples, see e.g. Fig. S2b below, has been subtracted.
\\
\\
\noindent The full-width-half-maximum (FWHM) of the resonances as a function of gate voltage ($V_g$) described by Eq.~22 provides a relation between temperature and lever arm:

\begin{align}
\Delta V_{g,FWHM}& = \frac{k_BT}{e \alpha}(ln(3+\sqrt{8})-ln(3-\sqrt{8})) \approx 3.53 \frac{k_BT}{e \alpha}
\end{align}
\\
\noindent The gate dependency is the differential of the fermi function, as expected. Using Eq. 7 and 22, the maximum phase signal is given by:
\begin{align}
|\Delta \Phi_{max}| = \frac{2Q}{C_{\Sigma}} \frac{e^2 \alpha^2}{4 k_B T}
\end{align}
\\
From these relations we can determine the temperature and lever arm as:
\\
\begin{equation}
T  = \frac{2 C_{\Sigma} |\Delta \Phi_{max}| \Delta V_{g,FWHM}^{2}}{(3.53)^{2} k_B Q} \quad \quad \quad \quad \alpha = \frac{2 C_{\Sigma} |\Delta \Phi_{max}| \Delta V_{g,FWHM}}{3.53 e Q }
\end{equation}
\\
\noindent We are thus able to determine the temperature from measurements of the peak width and height directly which does not require a separate measurement of the lever arm. The signal strength is maximized by improving the lever arm and quality factor. As described in the main text, the devices can be used as a primary thermometer, the limits of which are set by the tunnel coupling $\gamma$ and the 0D level spacing and charging energies of the nanoparticles.\\

\begin{figure*}[t]
\includegraphics[width=150mm]{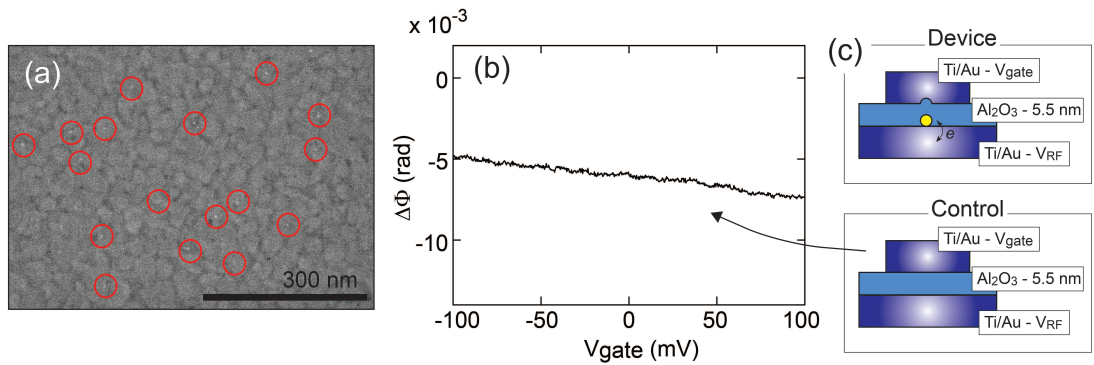}
\caption{\label{FigS2} \textbf{(a)} Scanning electron micrograph of a device \textit{with} Au nanoparticles as indicated by the red circles. \textbf{(b)} Phase data from a control device without nanoparticles for large 5x5 $\mu$m gate area. The vertical scale bar (phase) is set to allow for a direct comparison with Fig.~3b in the main text. \textbf{(c)} Schematic of the nanoparticle (top) and control (bottom) devices.}
\end{figure*}

\noindent To ensure that we measured the nanoparticles and not some other effect, e.g. charge traps in the oxides, control samples were prepared. When nanoparticles are deposited, a rough indication of the average density is obtained from scanning electron micrographs, see Fig.~S2a. For the control devices, exactly the same fabrication procedure as outlined in the Methods section in the main text was carried out, with the exception that no nanoparticles were deposited onto the substrates.  All other variables were kept the same, and the various control devices measured. An example is shown in Fig. S2b, for which indeed no signal is observed apart from a slowly decreasing background seen in all devices - leading to the conclusion that the nanoparticles are indeed responsible for resonances in the data. Samples \textit{with} Au nanoparticles but with considerable thicker gate oxides (> 10 nm Al$_2$O$_3$) also did not show a response at 400 mK (not shown) as expected since for these devices the lever arm will be much smaller and thus the measured signal strength, as expressed by Eq. 24.

\end{document}